\documentclass[12pt]{article}

\usepackage{amssymb,amsmath}

\usepackage{graphicx}
\DeclareGraphicsExtensions{.pdf,.png,.jpg}

\catcode`\@=11 \@addtoreset{equation}{section}\catcode`\@=12

\begin{document}

 \begin{center}

 \large \bf Stable exponential cosmological
solutions with three factor spaces in   $(1+ 3 + 3 +k)$-dimensional  
Einstein-Gauss-Bonnet model  with a $\Lambda$-term
  \end{center}

 \vspace{0.3truecm}

 \begin{center}

  K. K. Ernazarov

\vspace{0.3truecm}

 \it Institute of Gravitation and Cosmology,
 RUDN University, 6 Miklukho-Maklaya ul.,
 Moscow 117198, Russia

\end{center}

\begin{abstract} 

We consider a $(7 + k)$-dimensional  Einstein-Gauss-Bonnet model with the cosmological $\Lambda$-term.
A cosmological  model with three factor spaces of dimensions $3$, $3$ and $k$, $k > 2$ is considered. Exact stable solutions with three (non-coinciding) Hubble-like parameters in this model are obtained. Some examples of solutions (e.g. with zero variation of the effective gravitational constant $G$) are considered  in selected dimensions (for $k = 5, 6$).

\end{abstract}


\section{Introduction}

In this paper we consider a $D$-dimensional gravitational model
with Gauss-Bonnet term and cosmological term $\Lambda$, which
extend the model with cosmological $\Lambda$  term from ref. \cite{Deruelle}.

Attempts at the geometric description of dark energy led to the analysis of a number of possible extensions of general relativity. In addition to the "classical"  and scalar tensor theories, more complicated extensions have been developed, which are associated with the presence of torsion or more involved invariants of gravitational action. Among those theories that include a second-order derivative of the metric, one can distinguish the Einstein-Gauss-Bonnet gravitational theory, which has interesting properties.
The idea of these theories follows from the concept of cosmology on branes, which is based on string theory. In \cite{Lovelock_1} a new class of the gravity theory was proposed, called Lovelock gravity,while Gauss-Bonnet gravity is its particular case. The cosmology of Gauss-Bonnet gravity has been studied in detail in a number of papers and at present such exact analytical solutions as cosmological  (\cite{ElMakObOsFil} - \cite{IvKob-19-1}), centrally symmetric (generalization of Schwarzschild metric based on the Einstein-Gauss-Bonnet gravity) (\cite{R.Kon-1} - \cite{Torii}) and wormhole (\cite{Kanti} - \cite{Cu_Kon_Zhid}) solutions have been obtained.

\section{The setup}

In this  multidimensional Einstein-Gauss-Bonnet model, the metric is expressed in the following form:

\begin{equation}
 g= w du \otimes du + \sum_{i=1}^{n} e^{2\beta^i(t)}\epsilon_i  dy^i \otimes dy^i.
 \label{2.1A}
\end{equation}
on the manifold
\begin{equation}
   M = R  \times   M_1 \times \ldots \times M_n 
   \label{2.2A},
\end{equation}
where $ w=\pm1$, $\epsilon_i=\pm1$, $ i=1, ... , n$. The manifold M is defined as a set of one-dimensional manifolds $M_1, ... , M_n$.
In the open real set $R_*=(u_-, u_+) $, the functions $ \gamma(u) $ and $\beta^i(u)$  are smooth. The metric  (\ref{2.1A})  is cosmological for $w=-1$, $\epsilon_1=\epsilon_2= ...=\epsilon_n=1$  and for physical applications the manifolds $M_1$, $M_2$ and $M_3$  are equal to $\mathbb{R}$  and the other manifolds are considered as compact sets.

In this model, the action is expressed as
\begin{equation}
  S =  \int_{M} d^{D}z \sqrt{|g|} \{ \alpha_1 (R[g] - 2 \Lambda) +
              \alpha_2 {\cal L}_2[g] \},
 \label{2.3A}
\end{equation}
where $g=g_{MN}dz^M \otimes dz^N$ is the metric defined on the manifold M, $dimM=D$, $|g|=|det(g_{MN}| $ and

\begin{equation}
 {\cal L}_2 = R_{MNPQ} R^{MNPQ} - 4 R_{MN} R^{MN} +R^2
 \label{2.4A}
\end{equation}
is the standard Gauss-Bonnet term and $\alpha_1$, $\alpha_2$ are nonzero constants. Further we denote  $\alpha= \frac{\alpha_2}{\alpha_1}$. 


Recent astronomical observations and studies show that the Universe expands with acceleration instead of deceleration according to the scheme of the standard Friedmann model. Observations in the large-scale structure of the Universe show that visible matter and invisible dark matter can contribute only $ 31.7  \% $ of the total amount. The remaining $68.3  \%$   is dark energy, which causes an accelerated expansion of the Universe. Several interesting articles on Einstein-Gauss-Bonnet gravity have been published, which attempt to explain the problems of dark energy in cosmology (\cite{ChPavTop1} -  \cite{Myrzakulov_1}). It should be noted that in the Einstein-Gauss-Bonnet theory of gravity the expansion of the Universe may be explained without a cosmological term. This is the main feature of this theory of gravity.

Here we are dealing with the cosmological  solutions with diagonal metrics (of Bianchi-I-like type) governed by $n$ scale factors depending upon one variable, where $n>3$. Morever, we restrict ourselves by the solutions with exponental dependence of scale factors (with respect to synchronous variable $t$).
\begin{equation}
  a_i(t)  \sim exp(h^i t) 
   \label{2.5A},
\end{equation}
$i=1, ... ,n; $ $D = n+1$.

Recent astronomical observations show that our Universe in its present state expands with acceleration. Therefore, in order to describe the 3-dimensional exponential expansion of the Universe, we will assume that

\begin{equation}
  h^1=h^2=h^3=H>0
   \label{2.6A}.
\end{equation}

The integrand in  (\ref{2.3A}), when the metric  (\ref{2.1A}) is substituted, reads as follows
\begin{equation}
 \sqrt{|g|}\Bigl\{\alpha_1(R[g]-2\Lambda) +\alpha_2 {\cal L}_2[g]\Bigl\}=L+\frac{df}{du}
   \label{2.7A},
\end{equation}
where
\begin{equation}
 L=\alpha_1 L_1+\alpha_2 L_2
   \label{2.8A}.
\end{equation}

Here terms $L_1$ and $L_2$  are expressed in the following form \cite{IvKob-19-1}:
\begin{equation}
  L_1=(-w)e^{-\gamma+\gamma_0}G_{ij}\dot\beta^i\dot\beta^j - 2 \Lambda e^{\gamma+\gamma_0}
   \label{2.9A},
\end{equation}
\begin{equation}
 L_2=-\frac{1}{3}e^{-3\gamma+\gamma_0}G_{ijkl}\dot\beta^i\dot\beta^j\dot\beta^k\dot\beta^l
   \label{2.10A},
\end{equation}
where
\begin{equation}
 \gamma_0= \sum_{i=1}^n \beta^i
   \label{2.11A}.
\end{equation}

Here we use  a 2-linear symmetric form in the "mini-supermetric"  - 2 - metric of a pseudo-Euclidean signature:
\begin{equation}
<\upsilon_1, \upsilon_2> = G_{ij}\upsilon_1^i\upsilon_2^j
   \label{2.12A},
\end{equation}
where
\begin{equation}
 G_{ij}= \delta_{ij}-1 
   \label{2.13A},
\end{equation}
and  a 4-linear symmetric form -  Finslerian 4-metric:
\begin{equation}
<\upsilon_1, \upsilon_2, \upsilon_3, \upsilon_4> = G_{ijkl}\upsilon_1^i\upsilon_2^j\upsilon_3^k\upsilon_4^l
   \label{2.14A}
\end{equation}
with components 
\begin{equation}
 G_{ijkl}= (\delta_{ij}-1) (\delta_{ik}-1) (\delta_{il}-1) (\delta_{jk}-1) (\delta_{jl}-1) (\delta_{kl}-1) 
   \label{2.15A}.
\end{equation}

Here we denote $ \dot A=\frac{dA}{du} $, and the function $f(u)$ in  (\ref{2.7A})  is irrelevant for our consideration (see \cite{Iv-09}, \cite{Iv-10}).

The derivation of  (\ref{2.8A}) -  (\ref{2.10A})  is based on the following identities (\cite{Iv-09}, \cite{Iv-10}):
\begin{equation}
 G_{ij}\upsilon^i\upsilon^j= \sum_{i=1}^n (\upsilon^i)^2 - \Biggl(\sum_{i=1}^n \upsilon^i\Biggl)^2 
   \label{2.16A},
\end{equation}

\begin{eqnarray}
 G_{ijkl}\upsilon^i\upsilon^j\upsilon^k\upsilon^l= \Biggl( \sum_{i=1}^n \upsilon^i \Biggl)^4 - 6\Biggl(\sum_{i=1}^n \upsilon^i\Biggl)^2\sum_{j=1}^n (\upsilon^j)^2   \nonumber \\
+  3\Biggl(\sum_{i=1}^n (\upsilon^i)^2\Biggl)^2 + 8\Biggl(\sum_{i=1}^n \upsilon^i\Biggl) \sum_{j=1}^n (\upsilon^j)^3 - 6\sum_{i=1}^n (\upsilon^i)^4
   \label{2.17A}.
\end{eqnarray}

From the action  (\ref{2.3A}) we can get the following form of the equation of motion:

\begin{equation}
 \epsilon_{MN} = \alpha_1\epsilon_{MN}^{(1)} + \alpha_2\epsilon_{MN}^{(2)} = 0 
   \label{3.1A},
\end{equation}

where

\begin{equation}
 \epsilon_{MN}^{(1)} = R_{MN} - \frac{1}{2}Rg_{MN} + \Lambda 
   \label{3.2A},
\end{equation},

\begin{equation}
 \epsilon_{MN}^{(2)} = 2 \Biggl(R_{MPQS}R_N^{PQS} - 2R_{MP}R_N^P - 2R_{MPNQ}R^{PQ} + RR_{MN} \Biggl) - \frac{1}{2}{\cal L}_2 g_{MN}
   \label{3.3A}.
\end{equation}

Now we put $w= -1$,  and the equations of motion for the action (\ref{2.3A}) 
give us the set of  polynomial equations \cite{ErIvKob-16}
\begin{eqnarray}
  E = G_{ij} v^i v^j + 2 \Lambda
  - \alpha   G_{ijkl} v^i v^j v^k v^l = 0,  \label{3.4A} \\
   Y_i =  \left[ 2   G_{ij} v^j
    - \frac{4}{3} \alpha  G_{ijkl}  v^j v^k v^l \right] \sum_{i=1}^n v^i 
    - \frac{2}{3}   G_{ij} v^i v^j  +  \frac{8}{3} \Lambda = 0,
   \label{3.5A}
\end{eqnarray}

$i = 1,\ldots, n$, where  $\alpha = \alpha_2/\alpha_1$. 
For $n > 3$ we get a set of forth-order polynomial  equations.

We note that for $\Lambda =0$ and $n > 3$ the set of equations (\ref{3.4A}) 
and (\ref{3.5A}) has an isotropic solution $v^1 = \cdots = v^n = H$ only 
if $\alpha  < 0$ \cite{Iv-09,Iv-10}.
This solution was generalized in \cite{ChPavTop} to the case $\Lambda \neq 0$.

It was shown in \cite{Iv-09,Iv-10} that there are no more than
three different  numbers among  $v^1,\dots ,v^n$ when $\Lambda =0$. This is valid also
for  $\Lambda \neq 0$ if $\sum_{i = 1}^{n} v^i \neq 0$. 

\section{Exponential solutions with three factor spaces}

In this section we deal with a class  of solutions to the set of equations (\ref{3.4A}), 
(\ref{3.5A}) of the following form:
\begin{equation}
  \label{3.1}
   v =(\underbrace{H,H,H}_{``our'' \ space},\underbrace{\overbrace{h_1, \ldots, h_1}^{k_1}, \overbrace{h_2, \ldots, h_2}^{k_2}}_{internal \ space}),
\end{equation}
where $H$ is the Hubble-like parameter corresponding  
to an $3$-dimensional factor space,  $h_1$ is the Hubble-like parameter 
corresponding to an $k_1$-dimensional factor space with $k_1 > 1$ and $h_2$ ($h_2 \neq h_1$) is the Hubble-like parameter 
corresponding to an $k_2$-dimensional factor space with $k_2 > 1$. The first one is identified with  "our" $3d$ space while the next ones are considered as 
subspaces  of $( k_1 + k_2)$-dimensional internal space.
 
We assume  
\begin{equation}
  \label{3.2a}
   H > 0 
\end{equation}
for a description of an accelerated expansion of a
$3$-dimensional subspace (which may describe our Universe).

According to the ansatz (\ref{3.1}),  the first  $3$-dimensional factor space is expanding with the Hubble parameter $H >0$, while the $k_i$-dimensional internal  factor space  is contracting with the Hubble-like  parameter $h_i < 0$, where $i$ is 
either $1$ or $2$.

Now we consider the ansatz (\ref{3.1}) with three Hubble parameters $H$, $h_1$ and $h_2$ 
which obey the following restrictions:
   \begin{equation}
   S_1 = 3 H +  k_1h_1 + k_2 h_2 \neq 0, \quad  H \neq h_1, \quad  H \neq h_2, \quad  h_1 \neq h_2, \quad  k \neq 1.
   \label{3.3}
   \end{equation}

 It was proved in ref. \cite{ErIv-17-2} that  the set of $(n + 1)$ polynomial equations  
 (\ref{3.4A}), (\ref{3.5A}) under ansatz  
 (\ref{3.1}) and restrictions (\ref{3.3}) imposed  is reduced to a set  of three polynomial equations 
 (of fourth, second and first orders, respectively)
    
    \begin{eqnarray}
          E =0,   \label{3.4E} \\
          Q =  - \frac{1}{2 \alpha}, \label{3.4Q} \\
          L = H + h_1 + h_2 - S_1 = 0.  \label{3.4L}
     \end{eqnarray}
   where  $E$ is defined in (\ref{3.4A})  with ($v^1$,$v^2$,$v^3$) = (H, $h_1$, $h_2$) and 
   \begin{equation}
        Q = Q_{h_1 h_2} =  S_1^2 - S_2 - 2 S_1 (h_1 + h_2) + 2 (h_1^2 + h_1 h_2 + h_2^2),
                   \label{3.5}
        \end{equation}
$S_1$  is defined in (\ref{3.3})  and $ S_2 = 3 H^2 + k_1 (h_1)^2 + k_2 (h_2)^2$
     
 As it was proved in \cite{ErIv-17-2}  by using results of ref. \cite{Ivas-16-1} (see also \cite{Pavl-15}),  the
 exponential solutions with  $v$ from (\ref{3.1}) and $k_1 > 1$, $k_1 > 2$ are stable if and only if 
 \begin{equation}
           S_1 = 3 H +  k_1h_1 + k_2 h_2 = H + h_1 + h_2 > 0. 
                       \label{3.6}
 \end{equation}
Here we use the relation (\ref{3.4L}).

\subsection{Exact stable solutions in $(3+ 3+ k)$-dimensional case}

In this subsection, we present solutions to the set of equations of motion in  in the form:
\begin{equation}
v =(\underbrace{H,H,H}_{''our'' space},\underbrace{\overbrace{h_1, h_1, h_1}^3, \overbrace{h_2, \ldots, h_2}^{k_2}}_{internal \ space})
\label{6.1A}
\end{equation}

\noindent 
where $k_2 = k > 2$  and $H$ is the Hubble-like parameter that corresponds to the 3-dimensional “our” subspace and the $h_1$, $h_2$ are Hubble-like parameters that correspond to the internal subspaces of dimensions
$3$ and $k_2 >2$, respectively.
These solutions may be readily rewritten to the case
 \begin{equation}
v =(\underbrace{H,H,H}_{''our'' \ space},\underbrace{\overbrace{h_1, \ldots, h_1}^{k_1}, \overbrace{h_2, h_2, h_2}^3}_{internal \ space}),
\label{6.2A}
\end{equation}
where $k_1 >1$.

Our solutions must satisfy the following conditions:

A) $H>0$. This condition is necessary to describe the accelerated expansion of "our" 3-dimensional world, i.e., we assume that our 3-dimensional world corresponds to an expanding subspace in the multidimensional model. The remaining dimensions are considered as an internal subspace dimensions.

B) As it is well known, our multidimensional model is considered anisotropic, therefore expansion is carried out in some dimensions and the remaining dimensions are compressed. In this particular case, we believe that in "our" 3-dimensional world there is an expansion, and the rest of the internal dimensions either have a contraction, or an expansion in some dimensions and a contraction in the other internal dimensions. Therefore, the fulfillment of this condition is required:

\noindent B.1) ($h_1 < 0$, $h_2 < 0$) - a contraction in the internal subspace; \newline
B.2) ($h_1 < 0$, $h_2 > 0$) - a contraction in the internal $k_1$-dimensions and an  expansion in the internal $ k_2 $-dimensions; \newline
B.3) ($h_1 > 0$, $h_2 < 0$) - an expansion in the internal $k_1$-dimensions and a contraction in the internal $k_2$ -dimensions. 

We note that the solutions with $H>0$, $h_1>0$, $h_2 >0$ do note appear in our consideration due to 
relation (\ref{6.6A}). The solutions obeying B.2) are unstable.  

When fulfilling the above conditions, from (\ref{3.4Q}) we get following solutions in case $(m, k_1, k_2) = (3, 3, k_2)$:

\begin{equation}
h_1 = - \frac{1}{4}\Biggl((k_2 -1)h_2 \pm\sqrt{\frac{2}{\alpha} - (k_2 +3)(k_2 -1)h_2^2}\Biggl)
\label{6.3A}.
\end{equation}.

The substitution of (\ref{6.3A}) and $\lambda=\Lambda\alpha$ into relation (\ref{3.4E}) gives us the following expression 

\begin{equation}
h_2 = \frac{\sqrt{\frac{1}{\alpha}(k_2+1)(k_2-1)(k_2 + 3)(k_2 - 3)\Biggl((k_2 - 1)(k_2-3) \pm 2A}\Biggl)}{(k_2 - 1)(k_2 + 1)(k_2 - 3)(k_2 +3)}
\label{6.4A},
\end{equation}

\begin{equation}
h_2 = -\frac{\sqrt{\frac{1}{\alpha}(k_2+1)(k_2-1)(k_2 + 3)(k_2 - 3)\Biggl((k_2 - 1)(k_2-3) \pm 2A}\Biggl)}{(k_2 - 1)(k_2 + 1)(k_2 - 3)(k_2 + 3)}
\label{6.4ABN},
\end{equation}

here
\begin{equation}
A=\sqrt{(k_2 - 1)(k_2 - 3)\Biggl((1 - 4\lambda)k_2^2 + 2(1 -   8\lambda)k_2 + 3(1 - 4\lambda)\Biggl)}
\label{6.5A}.
\end{equation}.

\noindent Above we denote part of the expression by A to shorten the long formulas.

Thus, from the last relations we can see that all the Hubble - like parameters $h_1$ and $h_2$ of the internal space are uniquely determined by $k_2$ and $\lambda$. The Hubble-like parameters of  ''our'' 3-dimensional world is carried out using equation (\ref{3.4L}):

\begin{equation}
H = - \frac{2h_1 + (k_2 - 1)h_2}{2}
\label{6.6A}.
\end{equation}

The stable solutions are selected  by relation 
\begin{equation}
S_1 =  \frac{(3 - k_2)}{2} h_2 > 0.
\label{6.6B}
\end{equation}
For $k_2 > 2$ we have stable solutions for $h_2 < 0$ and unstable  - 
for $h_2 > 0$. The case $k_2 = 2$ will be considered in a separate publication.

\section{Examples}

\subsection{$k_2 =5$ and $\alpha > 0$}

Let us consider the case $ k_2 = 5$. From (\ref{6.4A}) we get four solutions: 

\begin{equation}
h_2 = \frac{1}{4\sqrt{3\alpha}}\sqrt{1 \pm \sqrt{19 - 96\lambda}}
\label{7.1A},
\end{equation}

\begin{equation}
h_2 = -\frac{1}{4\sqrt{3\alpha}}\sqrt{1 \pm \sqrt{19 - 96\lambda}}
\label{7.1ABK}
\end{equation}

\noindent and further, as our calculations show, each value of four solution of $h_2$ corresponds to two values of the solution $h_1$ (see (\ref{6.3A})) and one number of the solutions H (see (\ref{6.6A})). Therefore, we can find eight number of the set of real solutions. As our calculations show, four of them are unstable.
 Therefore, when we introduce the stability condition and the conditions A), B), the number of the set of stable real solutions reduce to three:

1) 
\begin{equation}
H = \frac{1}{4\sqrt{3\alpha}}\Biggl(\sqrt{1 + \sqrt{19 - 96\lambda}} + \sqrt{4 - 2\sqrt{19 - 96\lambda}}\Biggl)
\label{7.2A},
\end{equation}

\begin{equation}
h_1 = \frac{1}{4\sqrt{3\alpha}}\Biggl(\sqrt{1 + \sqrt{19 - 96\lambda}} - \sqrt{4 - 2\sqrt{19 - 96\lambda}}\Biggl)
\label{7.3A},
\end{equation}

\begin{equation}
h_2 =- \frac{1}{4\sqrt{3\alpha}}\sqrt{1 + \sqrt{19 - 96\lambda}}
\label{7.4A},
\end{equation}

\begin{equation}
S_1 = \frac{1}{4\sqrt{3\alpha}}\sqrt{1 + \sqrt{19 - 96\lambda}} > 0
\label{7.4AB}.
\end{equation}

The solutions $H > 0$, $h_1 < 0$ and $h_2 < 0 $ are occured in the interval of $\lambda$:

\begin{displaymath} 
\frac{3}{16} < \lambda < \frac{19}{96}
\end{displaymath} 

and the solutions $H > 0$, $h_1 > 0$ and $h_2 < 0 $ are existed in the interval of $\lambda$: 
\begin{displaymath} 
\frac{5}{32} < \lambda < \frac{3}{16}.
\end{displaymath} 

2)

\begin{equation}
H = \frac{1}{4\sqrt{3\alpha}}\Biggl(\sqrt{1 + \sqrt{19 - 96\lambda}} - \sqrt{4 - 2\sqrt{19 - 96\lambda}}\Biggl)
\label{7.5AB},
\end{equation}

\begin{equation}
h_1 = \frac{1}{4\sqrt{3\alpha}}\Biggl(\sqrt{1 + \sqrt{19 - 96\lambda}} + \sqrt{4 - 2\sqrt{19 - 96\lambda}}\Biggl)
\label{7.6AB},
\end{equation}

\begin{equation}
h_2 =- \frac{1}{4\sqrt{3\alpha}}\sqrt{1 + \sqrt{19 - 96\lambda}}
\label{7.7AB},
\end{equation}

\begin{equation}
S_1 = \frac{1}{4\sqrt{3\alpha}}\sqrt{1 + \sqrt{19 - 96\lambda}} > 0
\label{7.7AB_2}.
\end{equation}

The solutions $H > 0$, $h_1 > 0$ and $h_2 < 0 $ are occured in the interval of $\lambda$:

\begin{displaymath} 
\frac{5}{32} <  \lambda < \frac{3}{16}.
\end{displaymath} 

3)

\begin{equation}
H = \frac{1}{4\sqrt{3\alpha}}\Biggl(\sqrt{1 - \sqrt{19 - 96\lambda}} + \sqrt{4 + 2\sqrt{19 - 96\lambda}}\Biggl)
\label{7.8AC},
\end{equation}

\begin{equation}
h_1 = \frac{1}{4\sqrt{3\alpha}}\Biggl(\sqrt{1 - \sqrt{19 - 96\lambda}} - \sqrt{4 + 2\sqrt{19 - 96\lambda}}\Biggl)
\label{7.9AC},
\end{equation}

\begin{equation}
h_2 =- \frac{1}{4\sqrt{3\alpha}}\sqrt{1 - \sqrt{19 - 96\lambda}}
\label{7.10AC},
\end{equation}

\begin{equation}
S_1 = \frac{1}{4\sqrt{3\alpha}}\sqrt{1 - \sqrt{19 - 96\lambda}} > 0
\label{7.10AC_2}.
\end{equation}

The solutions $H > 0$, $h_1 < 0$ and $h_2 < 0 $ are occured in the interval of $\lambda$:

\begin{displaymath} 
\frac{3}{16} < \lambda < \frac{19}{96}.
\end{displaymath}

\subsection{$k_1=6$ and $\alpha > 0$}

In the set of dimensions $ (m, k_1, k_2 ) = ( 3, 6, 3 )$, solving the set of polynomial equations ( (\ref{3.4E}) - (\ref{3.4L})), one can obtain analogous formulas and expressions as in the set of dimensions $ (m, k_1, k_2 ) = ( 3, 3, 5 )$. In this set of dimensions from (\ref{3.4E}) we get four real solutions for $h_1$ and for each value of four solution of $h_1$ corresponds to two values of the solution $h_2$ (see (\ref{3.4Q})) and four numbers of the solutions H (see (\ref{3.4L})). Therefore, eight number of the set of real solutions are occured. As our calculations show, four of them are unstable.
 Therefore, when we introduce the stability condition and the conditions A), B), the number of the set of stable real solutions decreases up to three:

1) 
\begin{equation}
H = \frac{1}{12\sqrt{7\alpha}}\Biggl(\sqrt{25 + 10\sqrt{85 - 420\lambda}} + 3\sqrt{9 - 2\sqrt{85 - 420\lambda}}\Biggl)
\label{7.11AD},
\end{equation}

\begin{equation}
h_1 =- \frac{1}{\sqrt{315\alpha}}(\sqrt{5 + 2\sqrt{85 - 420\lambda}}
\label{7.12AD},
\end{equation}

\begin{equation}
h_2 = \frac{1}{12\sqrt{7\alpha}}\Biggl(\sqrt{25 + 10\sqrt{85 - 420\lambda}} - 3\sqrt{9 - 2\sqrt{85 - 420\lambda}}\Biggl)
\label{7.13AD},
\end{equation}

\begin{equation}
S_1 = \frac{1}{2\sqrt{35\alpha}}(\sqrt{5 + 2\sqrt{85 - 420\lambda}} > 0
\label{7.13ADC}.
\end{equation}

The solutions $H > 0$, $h_1 < 0$ and $h_2 < 0 $ are occured in the interval of $\lambda$:

\begin{displaymath} 
\frac{27}{140} < \lambda < \frac{17}{84}
\end{displaymath} 

and the solutions $H > 0$, $h_1 < 0$ and $h_2 > 0 $ are existed in the interval of $\lambda$: 
\begin{displaymath} 
\frac{37}{240} < \lambda < \frac{27}{140}.
\end{displaymath}

2)

\begin{equation}
H = \frac{1}{12\sqrt{7\alpha}}\Biggl(\sqrt{25 - 10\sqrt{85 - 420\lambda}} + 3\sqrt{9 + 2\sqrt{85 - 420\lambda}}\Biggl)
\label{7.14AD},
\end{equation}

\begin{equation}
h_1 =- \frac{1}{\sqrt{315\alpha}}(\sqrt{5 - 2\sqrt{85 - 420\lambda}}
\label{7.15AD},
\end{equation}

\begin{equation}
h_2 = \frac{1}{12\sqrt{7\alpha}}\Biggl(\sqrt{25 - 10\sqrt{85 - 420\lambda}} - 3\sqrt{9 + 2\sqrt{85 - 420\lambda}}\Biggl)
\label{7.16AD},
\end{equation}

\begin{equation}
S_1 = \frac{1}{2\sqrt{35\alpha}}(\sqrt{5 - 2\sqrt{85 - 420\lambda}} > 0
\label{7.16ADC}.
\end{equation}

The solutions $H > 0$, $h_1 < 0$ and $h_2 < 0 $ are occured in the interval of $\lambda$:

\begin{displaymath} 
\frac{3}{16} < \lambda < \frac{17}{84}.
\end{displaymath} 

3)

\begin{equation}
H = \frac{1}{12\sqrt{7\alpha}}\Biggl(\sqrt{25 + 10\sqrt{85 - 420\lambda}} - 3\sqrt{9 - 2\sqrt{85 - 420\lambda}}\Biggl)
\label{7.17AD},
\end{equation}

\begin{equation}
h_1 =- \frac{1}{\sqrt{315\alpha}}(\sqrt{5 + 2\sqrt{85 - 420\lambda}}
\label{7.18AD},
\end{equation}

\begin{equation}
h_2 = \frac{1}{12\sqrt{7\alpha}}\Biggl(\sqrt{25 + 10\sqrt{85 - 420\lambda}} + 3\sqrt{9 - 2\sqrt{85 - 420\lambda}}\Biggl)
\label{7.19AD},
\end{equation}

\begin{equation}
S_1 = \frac{1}{2\sqrt{35\alpha}}(\sqrt{5 + 2\sqrt{85 - 420\lambda}} > 0
\label{7.19ADC}.
\end{equation}

The solutions $H > 0$, $h_1 < 0$ and $h_2 > 0 $ are occured in the interval of $\lambda$:

\begin{displaymath} 
\frac{37}{240} < \lambda < \frac{27}{140}.
\end{displaymath}

\section{Stable solutions with zero variation of G}

In this multidimensional Einstein-Gauss-Bonnet model,  the solutions with zero variation of the effective gravitational G are of particular interest. Certain results are obtained in papers (\cite{ErIvKob-16} - \cite{Ivas-16-1}) on exact solutions with zero variation of the effective gravitational constant G. The condition of zero variation of the effective gravitational constant G in our case reads

\begin{equation}
k_1h_1 + k_2h_2 = 0
\label{8.1A}.
\end{equation}

As shown by our early studies \cite{ErIv-17-2}, in any dimension of the multidimensional model there are many exact solutions that include the exact solution with zero variation of the effective gravitational constant G. In this case, the cosmological term $\Lambda$ is determined by formula (3.25), (see \cite{ErIv-17-2}). This means that each set of dimensions and the cosmological term $ (m, k_1, k_2, \lambda)$ corresponds to a set of exact cosmological solutions, which includes a solution with zero variation of the effective gravitational constant G. In this case we obtained \cite{ErIv-17-2}:

\begin{eqnarray}
\lambda(m, k_1, k_2) = \Lambda\alpha = \frac{1}{8P^2} (m + k_1 + k_2 -3)\Biggl[\Biggl(k_1 + k_2)(k_1 + k_2 - 2\Biggl)m^3 \nonumber \\ 
+ \Biggl(k_1^3 + k_2^3 + 11 (k_1^2 k_2 + k_1 k_2^2) - 19 (k_1^2 + k_2^2)
- 22 k_1 k_2 + 18 (k_1 + k_2)\Biggl) m^2 - \nonumber \\
\Biggl(8((k_1^3 + k_2^3) - 63 (k_1 + k_2)^2 - 8 k_1^2 (k_1 - 11) k_2 \nonumber \\ 
- 8 k_2^2 (k_2 - 11) k_2) - 32 k_1^2 k_2^2 + 54 (k_1 + k_2)\Biggl) m\nonumber \\ 
- \Biggl( 9 (k_1^3 + k_2^3) + 45 (k_1^2 + k_2^2) - 54 (k_1 + k_2) + 8 (k_1^2 + k_2^2) k_1 k_2 \nonumber \\ 
- 16 (k_1 + k_2 -10) k_1^2 k_2^2 - 9 (21 k_1 + 21 k_2 - 26) k_1 k_2\Biggl)\Biggl],\nonumber \\ 
\label{8.2A}
\end{eqnarray}

where,

\begin{eqnarray}
P = P(m, k_1, k_2) = -(m+ k_1 + k_2 -3)\Biggl(m(k_1 + k_2 - 2) + \nonumber\\
k_1(2k_2 - 5) + k_2(2k_1 - 5) + 6\Biggl) \neq 0
\label{8.3A},
\end{eqnarray}

In the case under consideration, when $( m, k_1, k_2 ) = (3, 3, 5)$ the value of the cosmological term $\lambda$ is equal to
\begin{equation}
\lambda = \Lambda\alpha = \frac{681}{3872}
\label{8.4A}.
\end{equation}

Then for the set of dimensions and the cosmological term $ (m, k_1, k_2, \lambda) = (3, 3, 5, \frac{681}{3872}$), we obtain the following stable real solutions:

1) $H = \frac{5}{4}\frac{1}{\sqrt{11\alpha}}$, $h_1 = \frac{H}{5}$, $h_2 = - \frac{3}{5}H$;

2) $H = \frac{1}{4}\frac{1}{\sqrt{11\alpha}}$, $h_1 = 5H$, $h_2 = - 3H$.

Here the  solution 2) is stable \cite{ErIv-17-2}.

In the second case, when $( m, k_1, k_2 ) = (3, 6, 3)$ the value of the cosmological term $\lambda$ is equal to
\begin{equation}
\lambda = \Lambda\alpha = \frac{19}{108}
\label{8.5A},
\end{equation}

\noindent and for the set of dimensions and the cosmological term $ (m, k_1, k_2, \lambda) = (3, 6,3, \frac{19}{108})$, we obtain the following stable real solutions:

1) $H = \frac{2}{3}\frac{1}{\sqrt{3\alpha}}$, $h_1 = -\frac{H}{2}$, $h_2 = \frac{H}{4}$;

2) $H =\frac{1}{6} \frac{1}{\sqrt{3\alpha}}$, $h_1 = -2H$, $h_2 = 4H$.

In this case  the  solution 2) is stable \cite{ErIv-17-2}.

\section{Conclusions}

We have considered the $(7 + k)$-dimensional Einstein-Gauss-Bonnet (EGB) model
with the $\Lambda$-term. 
By using the ansatz with diagonal cosmological metrics, we have found 
for $D = 7 + k $, $\alpha = \alpha_2 / \alpha_1 > 0$ and certain $\lambda = \alpha \Lambda$ 
a class of exponential solutions with three Hubble-like parameters $H >0$, 
$h_1$, and $h_2$ corresponding to submanifolds of 
dimensions $m=3$, $k_1 = 3$ and $k_2 > 2$ (or $m=3$, $k_1 > 2$ and $k_2 = 3$), respectively. The obtained solutions are exact and stable. Two examples of solutions (for $(m,k_1,k_2) = (3,3,5), (3,6,3)$) are considered.
As we know, the stability plays a predominant role in study of exact cosmological solutions. 
Therefore, we assume that the obtained results will be used in our next research.

{\bf Acknowledgments}
The publication was prepared with the support of the ''RUDN University
Program 5-100''. It was also partially supported by the Russian Foundation
for Basic Research, grant Nr. 19-02-00346.


\end{document}